\documentclass{jjap3}

\title{ Effect of Co Impurities on Superconductivity of FeSe$_{0.4}$Te$_{0.6}$ single crystals}
\author{Fuyuki  NABESHIMA$^{1}$, Yusuke KOBAYASHI, Yoshinori IMAI$^{1}$, Ichiro TSUKADA$^{1,2}$, and Atsutaka MAEDA$^{1}$\thanks{E-mail address: cmaeda@mail.ecc.u-tokyo.ac.jp}}
\inst{Department of Basic Sciences, University of Tokyo, Komaba, Meguro, Tokyo 153-8902, Japan \\
$^{1}$Transformative Reserch Project on Iron Pnictides (TRIP), Japan Science and Technology Agency, Bancho, Chiyoda, Tokyo 102-0075, Japan\\
$^{2}$Central Research Institute of Electric Power Industry, Yokosuka, Kanagawa 240-0196, Japan}
\abst{
The effect of Co doping on supercoductivity of FeSe$_{0.4}$Te$_{0.6}$ single crystals is investigated.  The superconducting transition temperature decreases linearly for Co doping with the rate -0.75 K/(Co \%).
On the other hand, the increase of the residual resistivity is less than 50 $\mu\Omega$cm for 4 \% Co doping.
These data are consistent with the interband scattering mechanism of superconductivity with the sign change ($s_\pm$ symmetry).
}

\begin{document}

\maketitle

\section{Introduction}
Fe based superconductors\cite{Kamihara,Review} have attracted much attention of condensed-matter physicists and chemists, since Fe, which is ``the representative" of magnetic atoms, occupies the most essential part of the crystal structure and plays the dominant role for 50 K class superconductivity to emerge.
Soon, it turned out that many Fe bands contribute to superconductivity both theoretically\cite{band} and experimentally\cite{ARPES1,ARPES2}, and the mechanism of superconductivity for such a novel type of multiply gapped superconductors is of central importance.
Mechanisms based on spin fluctuation, where interband anti-ferromagnetic scattering between hole bands around $\Gamma$ point and electron bands around M point plays an essential role, favor the condensate wave function with the so-called $s_\pm$ symmetry\cite{Kuroki,Mazin}.
On the other hand, mechanisms based on the orbital fluctuation favor the s$_{++}$ symmetry with the same sign of the order parameter for both hole bands and electron bands\cite{Kontani,Ono}.
The absence of the coherence enhancement in the temperature dependence of the spin relaxation rate\cite{NMR1,Review} and microwave conductivity\cite{Hashimoto1,Hashimoto2,Imai,Takahashi} in many materials suggests the sign changing $s_\pm$ wave.
More detailed study of the coherence factor by STM in FeSe$_{1-x}$Te$_x$ (the so-called 11 material)\cite{Hanaguri} also suggests the $s_{\pm}$ wave definitely.
However, neutron scattering experiment\cite{neutron} and its interpretation\cite{Onarin}, and some of the investigation of the impurity effect\cite{Sato} suggests the $s_{++}$ wave.
There have been many studies on disorder effect\cite{Guo,Cheng,Kim1,Nakajima,Li}, whose results are complicated and there is no consensus on the possible pairing mechanisms, so far as the pair breaking effect is concerned.
For instance, the Co doping study in the 1111 material showed that $T_c$ decreases by 12 K by the substitution of Co by 5 \%.
According to the theoretical estimation\cite{Onari}, only the doping of impurity by by1 \% is expected to destroy superconductivity completely when the impurity potential is strong in the case of the $s_{\pm}$ wave.
This seems to be in strong contradiction to the experimental result.
However, even in the $s_{\pm}$ scenario, the decreasing rate of $T_c$ can become comparable to the experimentally observed number, when the strength of the impurity potential is very weak.
To check the strength of the impurity potential, one of the most important measurable quantities is the residual resistivity.
Thus, for the discussion of the symmetry of the condensate wave function by the pair-breaking effect, it is crucially important to discuss the decrease of $T_c$ and the increase of the residual resistivity, simultaneously.  In terms of this, in the experiments in polycrystalline samples\cite{Sato,Guo,Cheng}, large additional resistivity at the grain boundary masks the intrinsic behavior of the residual resistivity.
In addition, the effect of impurities in the 1111 material, where 5 elements exists in the sample other than the intentionally doped impurities might be very complicated.

Recently, studies of the effect of impurity (disorder) using single crystals of the 122 material came out\cite{Kim1,Nakajima,Li}.
Although all of these studies observed a considerable decrease of $T_c$ with small amount of disorder, conclusions are different among these three papers.
In addition, recent theoretical re-investigation of the effect of disorder suggest that there is a transition from the $s_{\pm}$ state to the $s_{++}$ state with increasing disorder.
Thus, there is almost no consensus on the effect of disorder in Fe-based superconductors, both experimentally and theoretically.
Thus, we focus on the simplest material among Fe-based superconductor family, the so-called 11 type chalcogenide, Fe(Se,Te)\cite{Wu}.
This material has $T_c$ of 14-15 K in a bulk form.
By making good films\cite{Imai2}, $T_c$ raises up to 16 K\cite{Tsukada} to 20 K\cite{Belingeli}.
We have prepared a series of single crystals of Fe$_{1-x}$Co$_{x}$Se$_{0.4}$Te$_{0.6}$ with $x =$ 0, 1, 2, 4 \%, and investigate how superconductivity changes with $x$, together with the residual resistivity.
Our results are consistent with the $s_\pm$ pairing, thus, suggest that the spin-fluctuation mechanism with the interband scattering is the appropriate description of superconductivity in Fe chalcogenides.

\section{Experiments}

Prior to the single crystal study, we investigated the series of substitution studies in the polycrystalline samples of the Fe 11 materials for Cr, Mn, Co, Ni, Ru, Pd, Pt, Ir, and found that Co did substitute the Fe site.
Thus, we chose Co as the representative of the substitution study. 

Single crystals of Fe$_{1-x}$Co$_{x}$Se$_{0.4}$Te$_{0.6}$ with $x=$ 0, 1, 2, 4 \% were prepared by the method described elsewhere\cite{Takahashi}.
Composition analysis by the EDX method revealed that the samples with the nominal composition of Fe:Se:Te:Co = 1:0.4:0.6:0.02 shows the actual ratio of Fe:Se:Te:Co = 1:0.32:0.64:0.03.  Considering the possible measurement errors in the EDX measurement, the result shows that Co substitutes the Fe site with almost the same amount as the nominal one.
One delicate issue is that the material has two different Fe sites, and the above EDX result does not give us any detailed information on these.
Exact experimental estimation of the amount of Fe atoms for each sites very difficult.
In general, however, if the change of the distribution of Fe atoms among these two different sites affects $T_c$, it should accompany a large change in resistibity (both the magnitude an the temperature dependence), which was not observed in our experiment, as will be shown below.
Thus, we believe that the change of the Fe distribution between the two different sites is negligible, or at least, is small so that the main conclusion of the paper is not affected at all.

Lattice constants were measured by an X-ray diffractometer.
Dc resistivity was measured by the four-probe method.
To obtain good reproducibility, the use of Au paste (Tokuriki \# 8560) was found to be crucial.
Superconductivity was also checked by the measurement of dc magnetization using a SQUID magnetometer.

\section{Experimental Results}

Figure 1 shows the a-axis and the c-axis lattice constants as a function of Co doping up to 4 \%.
Very slight change of a and c parameters suggests that the possible change of carrier by the introduction of Co is very small (almost negligible)\cite{Yang}.

\begin{figure}
\begin{center}
\includegraphics[width=0.6\linewidth]{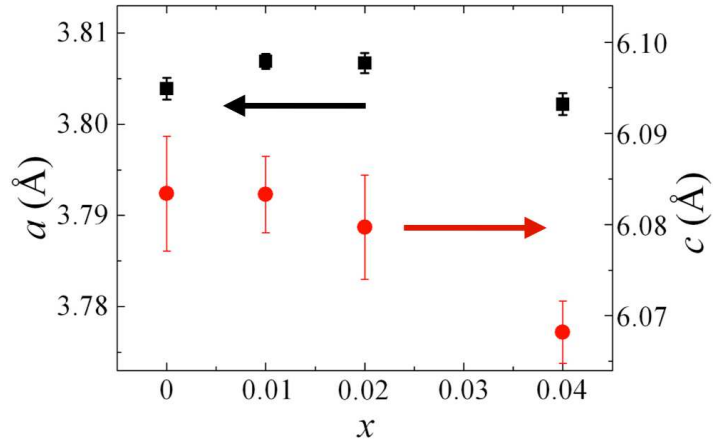}
\end{center}
\caption{Lattice constants of Fe$_{1-x}$Co$_{x}$Se$_{0.4}$Te$_{0.6}$ with $x=$ 0, 1, 2, and 4 \%.}
\label{f1}
\end{figure}

Figure 2 shows the temperature dependence of the dc susceptibility of the samples with $x=$ 0, 1, 2, 4 \%, respectively.
$T_c$ changes from 14.0 K for the $x=$0 sample to 11.0 K for the $x=$ 4 \% sample, monotonically with increasing Co content.
The transition width, $\Delta T_c$, defined as the difference of the temperatures for 10 \% and 90 \% of the diamagnetism at low temperatures is 0.8$\sim$1.0 K for all samples, showing that our single crystals are one of the best-quality crystals among the ones currently available for the same materials.

\begin{figure}
\begin{center}
\includegraphics[width=0.7\linewidth]{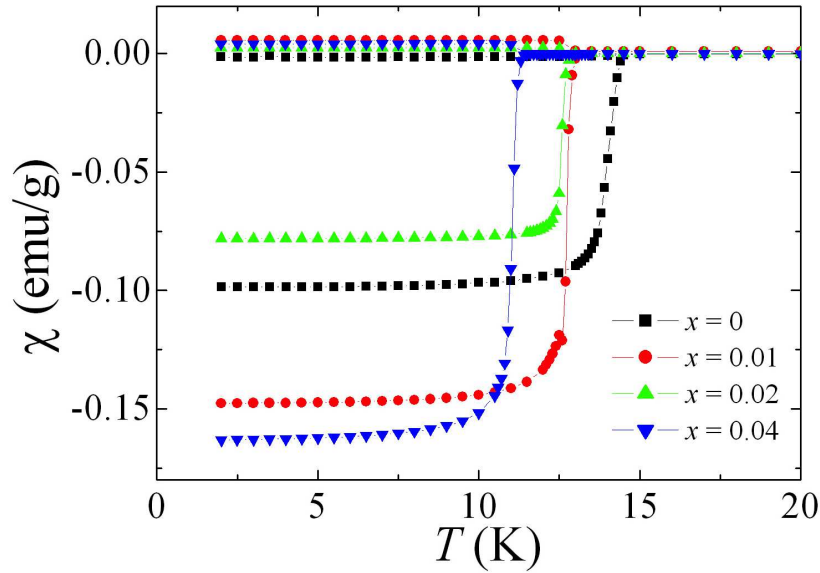}
\end{center}
\caption{Temperature dependence of dc magnetic susceptibility of Fe$_{1-x}$Co$_{x}$Se$_{0.4}$Te$_{0.6}$ with $x=$ 0, 1, 2, 4 \%.
Both zero-field cooled data and field cooled data are shown.}
\label{f2}
\end{figure}

Fiure 3 shows the temperature dependence of dc resitivity of 4 different pieces of the $x=$ 1 \% sample.
With this figure, we show that the reproducibility of the resistivity data (the temperature dependence, the magnitude, and the $T_c$ value, {\it etc.}) is rather good.
We define the residual resistivity of this material, as the extrapolated value of the low temperature linear part, as is shown in the inset of Fig. 3.

\begin{figure}
\begin{center}
\includegraphics[width=0.7\linewidth]{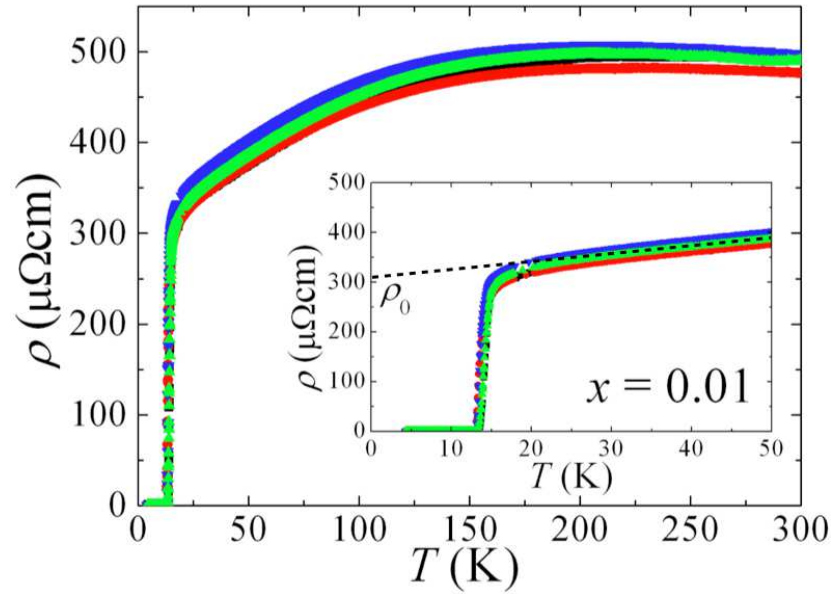}
\end{center}
\caption{Temperature dependence of dc resistivity of 4
pieces of Fe$_{1-x}$Co$_{x}$Se$_{0.4}$Te$_{0.6}$ with $x=$ 1 \%.
Dashed line is the extrapolation of the linear part to define the residual resistivity.}
\label{f3}
\end{figure}

Figure 4 shows the temperature dependence of dc resistivity of samples with different Co concentrations ($x=$ 0, 1, 2, 4 \%, respectively).
Within this range of Co concentration, the temperature dependence of resistivity is almost the same.
This suggests that Co behaves as nonmagnetic impurities.
With further doping of Co, the resistivity shows an upturn at the lowest temperatures.
Thus, we will discuss the Co doping effect within the Co concentration range shown in Fig. 4.

\begin{figure}
\begin{center}
\includegraphics[width=0.7\linewidth]{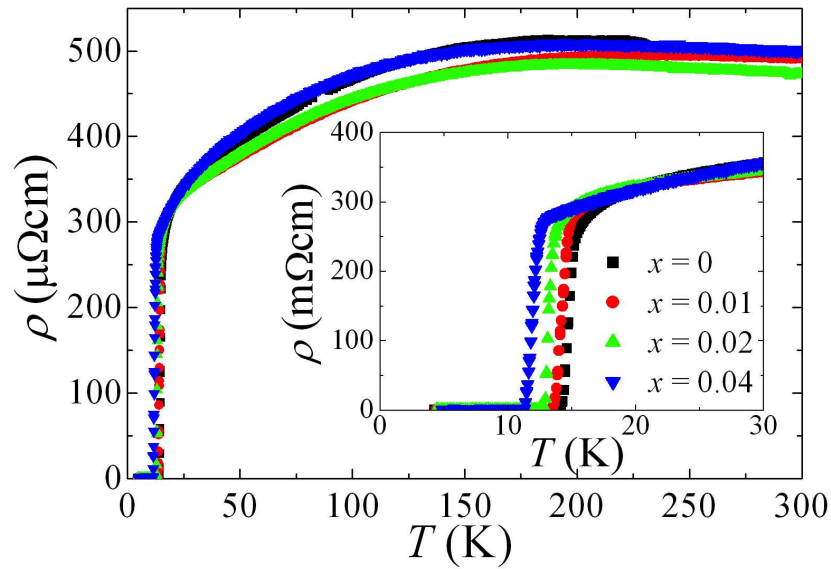}
\end{center}
\caption{Temperature dependence of dc resistivity of Fe$_{1-x}$Co$_{x}$Se$_{0.4}$Te$_{0.6}$ with x=0, 1, 2, 4 \%.
The inset shows the enlarged plot around $T_c$.}
\label{f4}
\end{figure}

From the data of Figs. 2 and 4, we plotted the superconducting transition temperature, $T_c$. as a function of Co content in Fig. 5.
$T_c$ was found to decrease almost linearly, with the rate, d$T_c/$d$x=$ -0.75 K/(Co \%).
We also obtained almost the same result in polycrystals, so far as the decrease of $T_c$ is concerned.

\begin{figure}
\begin{center}
\includegraphics[width=0.5\linewidth]{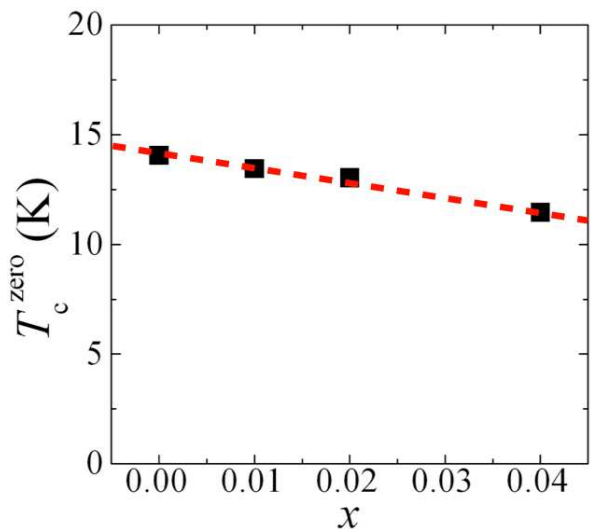}
\end{center}
\caption{$T_c$ as a function of Co concentration.
The dashed line is the straight line fit to the data.}
\label{f5}
\end{figure}

Figure 6 shows the residual resistivity, $\rho_0$, as a function of Co concentration.
For each concentration, error bars come from the scattering of the measured resistivity value among 4 pieces of crystal with the same Co contents.
From this data, the maximum possible increase of $\rho_0$ is found to be 50 $\mu\Omega$cm for 4 \% Co doping (12.5 $\mu\Omega$cm/(Co \%)).

\begin{figure}
\begin{center}
\includegraphics[width=0.5\linewidth]{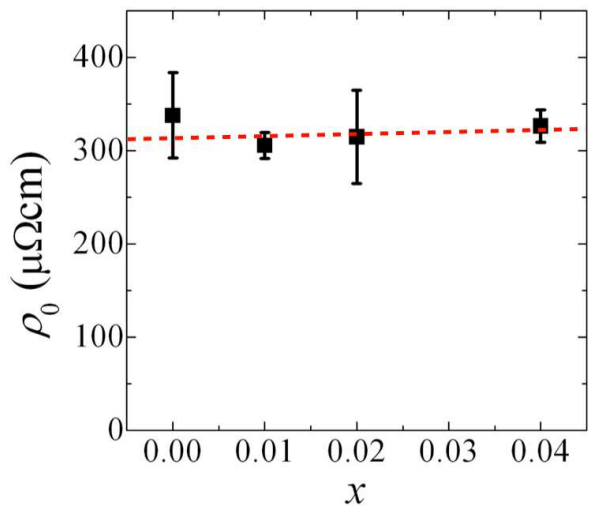}
\end{center}
\caption{The residual resistivity, $\rho_0$, as a function of Co concentration.
The dashed line is the theoretical expectation for the $s_\pm$ pairing. 
See the text for details.}
\label{f6}
\end{figure}

\section{Discussion}

The above results show that $T_c$ decreases with Co doping.
It is well established that superconductivity is robust for nonmagnetic disorder for conventional $s$-wave superconductivity\cite{Anderson}.
The magnetic disorder alone affects superconductivity, leading to the decrease of $T_c$\cite{AG}.
However, as was already mentioned, Co is considered to behave as nonmagnetic disorder in these materials.
Thus we do not expect the decrease of $T_c$ by 3 K for only 4 \% Co doping by the pair breaking effect.
Even in the case of nonmagnetic disorder, a slight decrease of $T_c$ is expected for two dimensional superconductors when one takes account of the weak localization effect\cite{WL}.
In this case, however, there must be a change in the temperature dependence of resistivity at low temperatures.
This is not the case for our present data.
Essential physics should be the same even for multiband superconductors with the same sign of the superconducting gaps for all Fermi surfaces.
It should be also added that the two dimensional nature is rather weak in the 11 material\cite{Hagiwara}.
Thus, in any cases, our results are hard to be understood in terms of the so-called $s_{++}$ wave pairing, unless the carrier concentration changes largely even by the very small Co substitution, which might cause the decrease of $T_c$.
However, for $T_c$ to be decreased by the change in carrier concentration in Fe(Se,Te) system, the large change (increase) of the resistivity is accompanied\cite{Noji}.
Therefore, it is unlikely that the change of $T_c$ shown here is caused by the change of carrier concentration.

An alternative possibility is the pair-breaking effect in superconductors with $s_{\pm}$ pairing.
In the $s_{\pm}$ pairing, the interband scattering is strongly affected by the presence of disorder, leading to very rapid decrease of $T_c$ with even very small amount of disorder\cite{Onari}.
Even the nonmagnetic disorder plays the same role as the magnetic impurities in conventional (or $s_{++}$) superconductors\cite{Chubkov}.
We check this possibility quantitatively, using our resistivity data.
According to Kontani\cite{Kontanip}, when one takes the effective mass ration as $m^*/m=$10, the ration of the change in $T_c$ to the change in the residual resistivity, $\Delta T_c/\Delta\rho_0$ is about $-0.3$ K/$\mu\Omega$cm.  Thus, we expected the increase of the residual resistivity to the Co concentration is 3$\mu\Omega$cm/Co \%, which we plotted in Fig. 6 as a straight line.
Our results do not contradict with the expectation for the $s_{\pm}$ pairing, in contrary to the argument in ref.\cite{Sato}.
The reason for the difference in the conclusion is that our data ara obtained in single crystals, and are free from the additional resistivity generated at grain boundaries characteristic of ploycrystals.

Indeed, the microwave conductivity data in the same material\cite{Takahashi,Kim} showed that low-temperature penetration depth shows the $T^2$ behavior, meaning that there is a large number of disorder even in Co free material.  This is also consistent with the $s_\pm$ scenario.

\section{Conclusion}

The effect of Co doping on superconductivity of FeSe$_{0.4}$Te$_{0.6}$ single crystals is investigated.  The superconducting transition temperature decreases linearly for Co doping with the rate -0.75 K/(Co \%).
On the other hand, the residual resistivity increase is less than 50 $\mu\Omega$cm for 4 \% Co doping.
These data are consistent with the interband scattering mechanism of superconductivity with the sign change ($s_\pm$ symmetry).

\acknowledgment
We thank Yoko Kiuchi, and staff of the Materials Design and Characterization Laboratory in ISSP for experimental support,
Jun-ichi Shimoyama for providing us the EDX apparatus, Hiroshi Kontani for fruitful discussions.



\end{document}